\documentclass[revtex4,apj,iop]{emulateapj}
\usepackage{natbib}
\usepackage{epsfig}
\usepackage{xcolor}
\usepackage{graphicx}
\usepackage{hyperref}

\newcommand{\ha}{H$\alpha$}
\newcommand{\msun}{\mbox{\rm M$_{\sun}$}}

\newcommand{\um}{$\mu$m}
\newcommand{\mic}{$\mu$m}
\newcommand{\hi}{\mbox{\ion{H}{1}}}
\newcommand{\hii}{\mbox{\ion{H}{2}}}
\newcommand{\msunyr}{\mbox{$\rm M_{\sun}\,yr^{-1}$}}
\newcommand{\spitzer}{\mbox {\it Spitzer}}
\newcommand{\galex}{\mbox {\it GALEX}}
\newcommand{\wise}{\mbox {\it WISE}}
\newcommand{\IRAS}{{\it IRAS}}

\newcommand \acorr {$a_{\rm corr}$}

\begin{document}

\title{STAR FORMATION HISTORIES ACROSS THE INTERACTING\\
       GALAXY NGC~6872, THE LARGEST-KNOWN SPIRAL}

\shorttitle{SFH ACROSS THE LARGEST-KNOWN SPIRAL GALAXY}

\author{Rafael~T.~Eufrasio\altaffilmark{1,2~\dag}, Eli~Dwek\altaffilmark{2}, Richard~G.~Arendt\altaffilmark{2,3}, Duilia~F.~de~Mello\altaffilmark{1,2,4}, Dimitri~A. Gadotti\altaffilmark{5}, Fernanda~Urrutia-Viscarra\altaffilmark{6}, Claudia~Mendes~de~Oliveira\altaffilmark{6}, \&  Dominic~J.~Benford\altaffilmark{2}}

\altaffiltext{1}{Physics Department, The Catholic University of America, Washington, DC 20064, USA}
\altaffiltext{2}{Observational Cosmology Laboratory, Code 665, NASA Goddard Space Flight Center, Greenbelt, MD 20771, USA}
\altaffiltext{3}{CRESST, University of Maryland-Baltimore County, Baltimore, MD 21250, USA}
\altaffiltext{4}{Johns Hopkins University, Baltimore, MD 21218, USA}
\altaffiltext{5}{European Southern Observatory, Santiago, Chile}
\altaffiltext{6}{Instituto de Astronomia, Geof\'isica e Ci\^encias Atmosf\'ericas, Universidade de S\~ao Paulo, Cidade Universit\'aria, 05508-090, S\~ao Paulo, Brazil}
\altaffiltext{\dag}{e-mail: rafael.t.eufrasio@nasa.gov}

%%%%%%%%%%%%%%%%%%%%%%%%%%%%%%%%%%%%%%%%%%%%%%%%%%%%%%%%%%%%%%%%%%%
\begin{abstract}

NGC~6872, hereafter the Condor, is a large spiral galaxy that is interacting with its closest companion, the S0 galaxy IC 4970. The extent of the Condor provides an opportunity for detailed investigation of  the impact of the interaction on the current star formation rate and its history across the  galaxy, on the age and spatial distribution of its stellar population, and on the mechanism that drive the star formation activity.  To address these issues we analyzed the far-ultraviolet (FUV) to near-infrared (near-IR) spectral energy distribution (SED) of 17, 10~kpc diameter, regions across the galaxy, and derived their star formation history, current star formation rate, and stellar population and mass.  We find that most of the star formation takes place in the extended arms,  with very little star formation in the central 5~kpc of the galaxy, in contrast to what was predicted from previous numerical simulations. There is a trend of increasing star formation activity with distance from the nucleus of the galaxy, and no evidence for a recent increase in the current star formation rate (SFR) due to the interaction. The nucleus itself shows no significant current star formation activity. The extent of the Condor also provides an opportunity to test the applicability of a single standard prescription for conversion of the FUV + IR (22 micron) intensities to a star formation rate for all regions. We find that the conversion factor differs from region to region, arising from regional differences in the stellar populations. 

\end{abstract}
%%%%%%%%%%%%%%%%%%%%%%%%%%%%%%%%%%%%%%%%%%%%%%%%%%%%%%%%%%%%%%%%%%%


%%%%%%%%%%%%%%%%%%%%%%%%%%%%%%%%%%%%%%%%%%%%%%%%%%%%%%%%%%%%%%%%%%%
\section{INTRODUCTION} \label{sec:intro}
%%%%%%%%%%%%%%%%%%%%%%%%%%%%%%%%%%%%%%%%%%%%%%%%%%%%%%%%%%%%%%%%%%%
The most accepted theory for the formation of galaxies is that they formed in a series of hierarchical mergers over cosmic time, which played an important role in their subsequent growth and morphological evolution \citep{SP1999,Lotz2004}. However, the detailed effects of the mergers and interactions on the subsequent evolution of the galaxies are still  uncertain. It is therefore of great importance to identify and study  local galaxies that are currently undergoing such interactions. Multi-wavelength observations are of paramount importance in reconstructing the star formation history (SFH) of interacting galaxies and in identifying the spectral signatures of their interactions. The far-ultraviolet (FUV) to mid-infrared (mid-IR) coverage of NGC~6872 and IC~4970 (hereafter the Condor galaxy and its companion, respectively) provides important information on their youngest star forming regions and total stellar mass, rendering this system a perfect laboratory for studying the effects of galaxy interactions on their spectral energy distributions (SEDs) \citep{Machacek2009}.

The Condor galaxy is a giant interacting spiral located in the dynamically young southern Pavo Group together with 12 other members \citep{Machacek2005}. We place it in this work at a distance of 65~Mpc \citep{Bastian2005}, and adjust all the previous literature values to this distance citing the original distances and measurements. The Condor has been the focus of several observational and theoretical studies in the past and has been suspected to be one of the largest-known spiral galaxies \citep{Block1979}. Using UV observations made with the {\it Galaxy Evolution Explorer} (\galex) satellite we show that the  arms of the Condor are longer than any other known spiral galaxy, measuring more than 160\,kpc (projected) from tip to tip. The physical extent of its arms is often attributed to the recent interaction of the Condor with its closest companion IC~4970, an S0 galaxy located 1.1\arcmin\ north of Condor's center. The Condor, however, has also interacted with an elliptical galaxy in the group, NGC~6876, located 8.7\arcmin\ to the SE of Condor's center. This interaction has formed the longest-known X-ray trail, extending from the Condor to NGC~6876 \citep{Machacek2005,Machacek2009}.

Numerical simulations of galaxy interactions also suggest the formation of tidal dwarf galaxies (TDGs) at the two far sides of the disk of the primary galaxy \citep{HM1995}. However, previous searches by \citet{Bastian2005} for these TDGs in the Condor have yielded no results. They found rich star clusters with ages between 1 and 100~Myr in the extended arms, using VLT/FORS1 in B, V and I. They also found that the eastern extended arm is forming stars at 2 times the rate of the western one and 5 times the rate of the main body, based on the U-band surface density.

The interaction between the Condor and its companion was first modeled by N-body simulations of \cite{Mihos1993}, assuming that the further elliptical galaxy NGC~6876 played little or no role in the dynamics of the pair and that the Condor was five times more massive than its closest companion. The model reproduced many morphological features of the interacting pair, including the formation of Condor's stellar bar and the extent of the optical tidal tails. However, the modeled spatial distribution of star formation did not reproduce the \ha\ observations. While the model displayed the majority of the star formation in the bar and in the central region of the galaxy, their \ha\ data supported star formation occurring mostly along the tidal arms.
At that time, the galaxy was thought to be very gas poor, since an \hi\ 21~cm non-detection by \cite{Reif1982} had assigned a 3-$\sigma$ upper limit of $4.6\times10^9~\msun$\footnote{originally $9.0\times10^{9}~\msun$ at 91.1~Mpc} of \hi\ for the whole galaxy. Based on the fact that the most intense \ha\ emitting regions coincided with regions with highest \ha\ velocities dispersions, \cite{Mihos1993} concluded the dominant mode of star formation in the Condor is collisionally-induced, and that, without the triggering by the interaction, the Condor would show little or no star formation at all.  They also noted the galaxy was cold compared to other interacting galaxies and that possibly a considerable fraction of its IR emission is not associated with the recent star formation rate (SFR).

Later \cite{HB1997} detected of a significantly larger reservoir with $2.15\times10^{10}~\msun$ of \hi\ gas for the whole system as well as $1.1\times10^9~\msun$ of hydrogen molecular gas\footnote{originally $1.82\times10^{10}~\msun$ of \hi\ and $9.6\times10^8~\msun$ of ${\rm H}_2$, respectively, at 59.8~Mpc} for the central 6.7~kpc of the Condor (43\arcsec\ in diameter), making the interacting pair one of the gas-richest known. More recently, \cite{HK07} presented higher resolution \hi\ interferometric data taken with ATCA (Australia Telescope Compact Array), from which they inferred the presence of only $1.75\times10^{10}~\msun$ of \hi\ gas\footnote{originally $1.54\times10^{9}~\msun$ total, with $1.41\times10^{9}~\msun$ for the Condor and $1.3\times10^{9}~\msun$ for the companion, at 61~Mpc}. Of these, they assigned $1.60\times10^{10}~\msun$ to the Condor, distributed in an extended rotating disk, and the remaining $1.5\times10^9~\msun$ to its companion. This means as much as $4.0\times10^9~\msun$ of missing \hi\ should be extended in scales larger than their expected size of the galaxy. No association of it with the X-ray trail between the Condor and the elliptical NGC~6876 was found. Their new \hi\ interferometric data show large concentrations of atomic gas in the spiral arms and a central area almost devoid of it. They also introduced a new N-body simulation including stars and gas, and also assumed the companion was five times less massive than the Condor. In the simulation, they were able to reproduce the thin arms as seen in the \hi\ and optical maps. Their analysis supports the scenario of a low-inclination prograde passage, now seen 130~Myr after closest approach.

Condor's large size enables a thorough analysis of the different sub-galactic regions that underwent star formation because of the interaction, and studies of the spatial variations in the resulting stellar and gas components of these regions, on scales of galactic proportions, of the order of $\sim10~{\rm pc}$. From this UV--to--mid-IR multi-wavelength dataset and evolutionary stellar population synthesis models, we derive stellar masses and star formation rates (SFRs) across the galaxy and compare them with the results of the simulations and previous analyses. We set out to (1) test if Condor's stellar bar was formed by the interaction, (2) test if star formation is concentrated in the bar and central areas, (3) test if the dominant mode of star formation is induced by the interaction, (4) obtain the ratio of stellar masses between the Condor and companion, and (5) check the use of UV+IR as a tracer of the SFR in the different regions.

The paper is organized as follows: In \S\ref{sec:data} we describe the data used in the analysis and the selected regions for detailed photometric analysis. In \S\ref{sec:SEDs} we derive the SEDs of regions across the galaxy. In \S\ref{sec:SFH} we describe our model to obtain the SFHs of the different regions.
In \S\ref{sec:results} we discuss the large scale trends derived from the SED fitting and compare the average SFR over the last 100~Myr to those derived using a UV+IR tracer of the SF activity. In \S\ref{sec:summary} we summarize the work.

%%%%%%%%%%%%%%%%%%%%%%%%%%%%%%%%%%%%%%%%%%%%%%%%%%%%%%%%%%%%%%%%%%%
\section{DATA \& METHOD} \label{sec:data}
\begin{figure*}
\begin{center}
  \includegraphics[width=1.\textwidth]{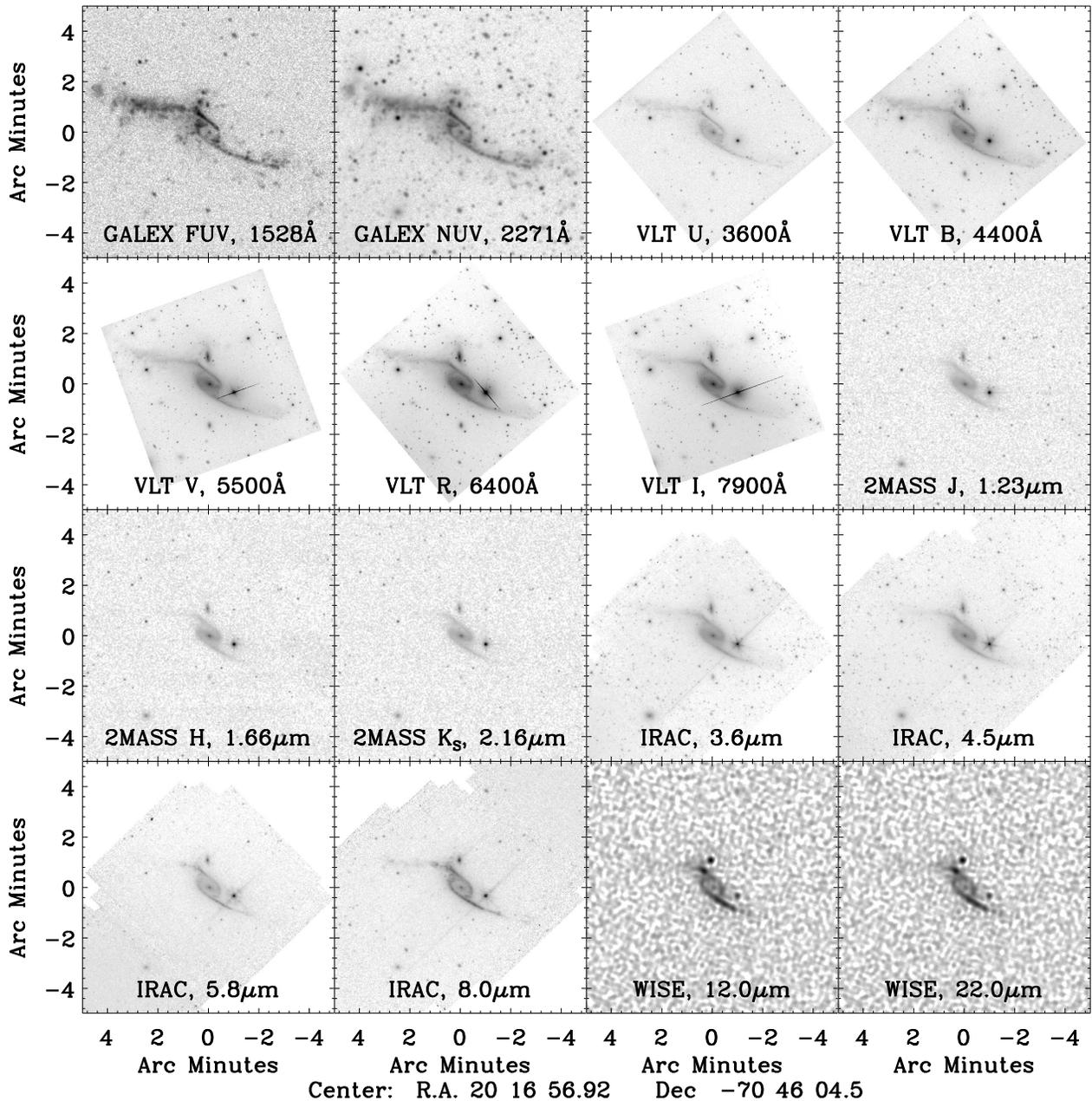}
  \caption{The broadband images of the Condor galaxy used here. The U-4.5~\um\ bands highlight progressively cooler stars in the arms, nucleus, and central bar of the galaxy. Increasing dust (PAH) emission at 5.8 and 8 \um\ highlights warm ISM in the arms and a ring that encloses the stellar bar. \galex\ UV images are dominated by star forming regions in the arms, and reveal hot stars in the TDG candidate at the tip of the NE arm.}
  \label{fig:tile}
\end{center}
\end{figure*}

%%%%%%%%%%%%%%%%%%%%%%%%%%%%%%%%%%%%%%%%%%%%%%%%%%%%%%%%%%%%%%%%%%%
For this work we made use of UV-to-IR archival data, summarized in Table~\ref{tab:data}. The table also lists all the instruments used in this paper, and related information. Figure~\ref{fig:tile} shows all the images in the same astrometric grid and scale.

%============================= Table 1 =============================
%===================================================================
\begin{deluxetable}{ccccc}[b]
\tabletypesize{\scriptsize}
\tablecaption{Multi-wavelength data set assembled for the Condor galaxy \label{tab:data}}
\tablewidth{0pt}
\tablehead{Telescope & $\lambda_0 \pm \Delta\lambda$\tablenotemark{a} [\um]& 
           FWHM\tablenotemark{b} & $f_{\rm Gal}({\lambda_0})$\tablenotemark{c} & $\sigma_{\rm cal}/C\tablenotemark{d}$
          }
\startdata
\galex\ FUV       & $ 0.15   \pm 0.02     $ & 4.3\arcsec & 2.676   & $15.0\%$ \\
\galex\ NUV       & $ 0.23   \pm 0.08     $ & 5.3\arcsec & 2.641   & $15.0\%$ \\
VLT U             & $ 0.36   \pm 0.05     $ & 1.5\arcsec & 1.583   & $ 5.0\%$ \\
VLT B             & $ 0.44   \pm 0.09     $ & 1.5\arcsec & 1.323   & $ 5.0\%$ \\
VLT V             & $ 0.55   \pm 0.11     $ & 0.5\arcsec & 1.000   & $ 5.0\%$ \\
VLT R             & $ 0.66   \pm 0.16     $ & 1.5\arcsec & 0.795   & $ 5.0\%$ \\
VLT I             & $ 0.77   \pm 0.14     $ & 0.5\arcsec & 0.551   & $ 5.0\%$ \\
2MASS J           & $ 1.23   \pm 0.21     $ & 3.0\arcsec & 0.260   & $10.0\%$ \\
2MASS H           & $ 1.66   \pm 0.26     $ & 3.0\arcsec & 0.165   & $10.0\%$ \\
2MASS K$_{\rm S}$ & $ 2.16   \pm 0.28     $ & 3.0\arcsec & 0.110   & $10.0\%$ \\
\wise\ (W1)       & $ 3.5\ \,\pm 0.6\ \,  $ & 2.8\arcsec & 0.050   & $ 2.4\%$ \\
\spitzer\ IRAC    & $ 3.6\ \,\pm 0.7\ \,  $ & 1.7\arcsec & 0.046   & $ 5.0\%$ \\
\spitzer\ IRAC    & $ 4.5\ \,\pm 1.0\ \,  $ & 1.7\arcsec & 0.029   & $ 5.0\%$ \\
\wise\ (W2)       & $ 4.6\ \,\pm 1.1\ \,  $ & 2.8\arcsec & 0.028   & $ 2.8\%$ \\
\spitzer\ IRAC    & $ 5.8\ \,\pm 1.4\ \,  $ & 1.9\arcsec & 0.020   & $ 5.0\%$ \\
\spitzer\ IRAC    & $ 8.0\ \,\pm 2.8\ \,  $ & 2.0\arcsec & 0.026   & $ 5.0\%$ \\
\wise\ (W3)       & $12.0\ \,\pm 6.3\ \ \ $ & 2.8\arcsec & \nodata & $ 4.5\%$ \\
\wise\ (W4)       & $22.0\ \,\pm 4.7\ \ \ $ & 7.2\arcsec & \nodata & $ 5.7\%$ \\
\enddata
\tablenotetext{a}{Central wavelength ($\lambda_0$) and FWHM of the filter transmission ($\Delta\lambda$).}
\tablenotetext{b}{FWHM spatial resolution of each map.}
\tablenotetext{c}{Galactic extinction curve normalized at the V band, i.e., $f_{\rm Gal}({\lambda_0}) = A_{\lambda}^{\rm Gal}/A_{\rm V}^{\rm Gal}$. For this work, we have used $A_{\rm V}^{\rm Gal} = 0.127$, according to the \citet{Schlafly2011} recalibration of the \cite{Schlegel1998} reddening maps. }
\tablenotetext{d}{Fractional calibration uncertainty.}
\end{deluxetable}

%============================ Figure 01 ============================
%===================================================================

Archival data from \galex\footnote{http://galex.stsci.edu/GalexView/}, Two Micron All Sky Survey (2MASS)\footnote{http://www.ipac.caltech.edu/2mass/releases/allsky/}, and the {\it Spitzer Space Telescope} (\spitzer)\footnote{http://irsa.ipac.caltech.edu/data/SPITZER/docs/\\ spitzerdataarchives/}\ were obtained from their respective archives and are being used as provided to the community. Extended source aperture corrections were applied to the \spitzer\ data as prescribed in the IRAC Instrument Handbook (V2.0.1)\footnote{http://irsa.ipac.caltech.edu/data/SPITZER/docs/irac/\\ iracinstrumenthandbook/IRAC\_Instrument\_Handbook.pdf}. Enhanced resolution data from the {\it Wide-Field Infrared Survey Explorer} (\wise)\footnote{http://wise2.ipac.caltech.edu/docs/release/allsky/}\ archive was processed according to \cite{Jarret2013}.

Optical U, B, V, R, and I images were retrieved from the European Southern Observatory archive\footnote{http://archive.eso.org/eso/eso\_archive\_main.html}.
The data were taken under photometric conditions with the FORS1 instrument at the Very Large Telescope (VLT) in the Atacama desert, in Chile. Standard bias subtraction and flat-fielding were performed. The images of each band were co-added and cleaned of cosmic rays. Flux calibration was determined by performing photometry on standard stars at different airmasses. All the observations, biases, flat-fields and standard stars frames were obtained from the VLT Science Archive.

\begin{figure}
%\begin{center}
  \hspace{-0.35in}
  \includegraphics[width=.55\textwidth]{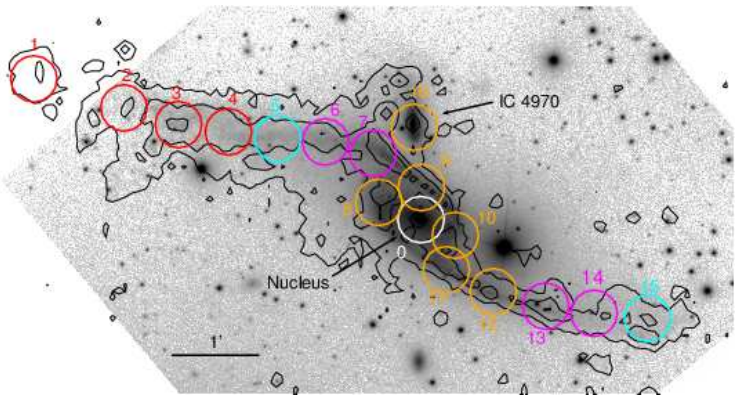}
  \caption{VLT R-band image of the Condor galaxy (NGC~6872). Contour levels are GALEX FUV at 0.002, 0.010, and 0.020 MJy/sr. Circles show the 17 regions analyzed (diameter of 32\arcsec). The optical image does not cover the entire field of the UV image. At the distance of 65~Mpc, the diameter of each region (32\arcsec) corresponds to a (projected) physical distance of 10~kpc. Color coding represents the trend of their SED colors as depicted in Fig.~\ref{fig:SEDs_norm} and described in \S~\ref{sec:SEDss}.}
  \label{fig:DSS_FUV_regions}
%\end{center}
\end{figure}

Figure~\ref{fig:DSS_FUV_regions} shows the 17 regions across the galaxy, for which we derived photometric measurements. They are 32\arcsec\ in diameter, which corresponds to a linear dimension of 10~kpc, for an assumed distance of 65~Mpc. The regions are not single star-forming regions, but large areas which include several \hii\ complexes. For scale comparison, the giant extragalactic \hii\ region in M\,33, known as NGC\,604, is 140 pc across \citep{MAMH2004} and the young (3\,Myr) Carina OB association in the Milky Way is also around 150~pc \citep{ZY2007}. Except for section \S\ref{sec:bar}, where we examine the morphology of the disk of the galaxy and properties of the stellar bar and Condor's nucleus in a smaller scale, all our analysis is done on these 10~kpc regions. Region 0 is Condor's central region and is comprised of its nucleus, most of its stellar bar and part of its disk. Region 16 corresponds to the companion galaxy, IC~4970.

%============================ Figure 02 ============================
%===================================================================

For each broadband image, the background was estimated by masking the galaxy and calculating the median of all the pixels outside the masked region, $\langle{F_{\nu}^{\rm bckg}}\rangle$. Then, a Gaussian was fitted to the histogram of background intensities (over $M$ pixels) to obtain its standard deviation $\sigma_M$. The background-subtracted flux $F_{\nu}$ of a given region is given by
%---------------- eq. 1
\begin{equation}
  F_{\nu} = C \left(\sum_{i=1}^{N} F_{\nu}^{\rm i} - N\times\langle{F_{\nu}^{\rm bckg}}\rangle\right),
\end{equation}
%----------------
where $F_{\nu}^{\rm i}$ is the $i$-th pixel in the region of interest (with $N$ total pixels) and $C$ is a calibration constant, whose associated uncertainty is $\sigma_{\rm cal}$. The fractional calibration uncertainties $\sigma_{\rm cal}/C$ for each image are given in Table~\ref{tab:data}. The uncertainty of a given background-subtracted flux is then
%---------------- eq 2
\begin{equation}
  \sigma_\nu = \sqrt{N\left(1+\frac{N}{M}\right)C^2\sigma_M^2 + \left(F_{\nu}\frac{\sigma_{\rm cal}}{C}\right)^2}.
\end{equation}
%----------------
The measured fluxes and their associated 1-$\sigma$ uncertainties are shown in Table \ref{tab:SEDs_1}.

\newpage

%============================= Table 1 =============================
%\clearpage
%\LongTables
%\begin{landscape}
\begin{deluxetable*}{cccccccccccc}
\tabletypesize{\scriptsize}
\tablecaption{Spectral Energy Distributions - Part 1 \label{tab:SEDs_1}}
\tablewidth{0pt}
\tablehead{Reg & R.A.      & Decl.      & FUV           & NUV           & U             & B              & V            & R            & I           & J          \\
               & (J2000)   & (J2000)    & (0.15\,\um)   & (0.23\,\um)   & (0.33\,\um)   & (0.44\,\um)    & (0.55\,\um)  & (0.64\,\um)  & (0.88\,\um) & (1.2\,\um)  }
\startdata
\\ 
 0\tablenotemark{a} & 304.23439 & -70.768263 & $0.07\pm0.01$ & $0.17\pm0.03$ & $3.10\pm0.16$ & $11.03\pm0.55$ & $23.8\pm1.2$ & $41.3\pm2.1$ & $68.0\pm3.4$ & $116\pm13$\\ \\ \hline \\
 1$^{\rm b}$ & 304.45822 & -70.741246 & $0.06\pm0.01$ & $0.07\pm0.01$ &    \nodata    &    \nodata     &   \nodata    &   \nodata    &   \nodata    & $    <  6$\\ \\ \hline \\
 2$^{\rm c}$ & 304.40606 & -70.746856 & $0.07\pm0.01$ & $0.08\pm0.01$ & $0.10\pm0.02$ & $ 0.22\pm0.05$ & $     < 0.1$ & $     < 0.1$ & $     < 0.3$ & $    <  6$\\
 3$^{\rm c}$ & 304.37443 & -70.750071 & $0.21\pm0.03$ & $0.25\pm0.04$ & $0.34\pm0.03$ & $ 0.63\pm0.06$ & $ 0.5\pm0.1$ & $ 0.6\pm0.1$ & $ 0.8\pm0.3$ & $    <  6$\\
 4$^{\rm c}$ & 304.34543 & -70.751468 & $0.15\pm0.02$ & $0.18\pm0.03$ & $0.31\pm0.03$ & $ 0.65\pm0.06$ & $ 0.7\pm0.1$ & $ 0.8\pm0.1$ & $ 1.2\pm0.3$ & $	   <  6$\\
 5$^{\rm c}$ & 304.31727 & -70.752861 & $0.15\pm0.02$ & $0.18\pm0.03$ & $0.33\pm0.03$ & $ 0.70\pm0.06$ & $ 0.9\pm0.1$ & $ 1.2\pm0.1$ & $ 1.9\pm0.3$ & $	   <  6$\\
 6$^{\rm c}$ & 304.28912 & -70.753406 & $0.13\pm0.02$ & $0.18\pm0.03$ & $0.42\pm0.03$ & $ 0.95\pm0.07$ & $ 1.3\pm0.2$ & $ 2.0\pm0.1$ & $ 3.0\pm0.3$ & $    <  6$\\
 7$^{\rm c}$ & 304.26181 & -70.755635 & $0.32\pm0.05$ & $0.46\pm0.07$ & $1.12\pm0.06$ & $ 2.50\pm0.13$ & $ 3.5\pm0.2$ & $ 5.8\pm0.3$ & $ 8.6\pm0.5$ & $ 13\pm 6$\\ \\ \hline \\
 8$^{\rm d}$ & 304.25747 & -70.764629 & $0.25\pm0.04$ & $0.39\pm0.06$ & $1.58\pm0.08$ & $ 4.54\pm0.23$ & $ 8.0\pm0.4$ & $14.0\pm0.7$ & $21.6\pm1.1$ & $ 36\pm 7$\\
 9$^{\rm d}$ & 304.23275 & -70.761794 & $0.15\pm0.02$ & $0.25\pm0.04$ & $1.15\pm0.06$ & $ 3.28\pm0.17$ & $ 6.3\pm0.3$ & $10.9\pm0.5$ & $18.1\pm0.9$ & $ 29\pm 6$\\
10$^{\rm d}$ & 304.21302 & -70.771052 & $0.08\pm0.01$ & $0.15\pm0.02$ & $1.16\pm0.06$ & $ 3.71\pm0.19$ & $ 7.6\pm0.4$ & $13.5\pm0.7$ & $23.6\pm1.2$ & $ 38\pm 7$\\ \\ \hline \\
11$^{\rm e}$ & 304.21894 & -70.777525 & $0.10\pm0.02$ & $0.18\pm0.03$ & $0.97\pm0.05$ & $ 2.72\pm0.14$ & $15.3\pm0.8$ & $ 8.4\pm0.4$ & $49.0\pm2.5$ & $ 23\pm 6$\\
12$^{\rm e}$ & 304.19156 & -70.781712 & $0.05\pm0.01$ & $0.08\pm0.01$ & $0.49\pm0.03$ & $ 1.46\pm0.09$ & $ 2.5\pm0.2$ & $ 4.6\pm0.2$ & $ 8.9\pm0.5$ & $ 11\pm 6$\\
13$^{\rm e}$ & 304.16162 & -70.784766 & $0.09\pm0.01$ & $0.11\pm0.02$ & $0.47\pm0.03$ & $ 1.10\pm0.07$ & $ 1.7\pm0.2$ & $ 6.0\pm0.3$ & $ 4.1\pm0.3$ & $    <  6$\\
14$^{\rm e}$ & 304.13341 & -70.786130 & $0.04\pm0.01$ & $0.05\pm0.01$ & $0.15\pm0.02$ & $ 0.38\pm0.05$ & $ 0.6\pm0.1$ & $ 0.9\pm0.1$ & $ 1.8\pm0.3$ & $    <  6$\\
15$^{\rm e}$ & 304.10263 & -70.787205 & $0.07\pm0.01$ & $0.09\pm0.01$ & $0.18\pm0.02$ & $ 0.37\pm0.05$ & $ 0.4\pm0.1$ & $ 0.5\pm0.1$ & $ 1.0\pm0.3$ & $    <  6$\\ \\ \hline \\
16$^{\rm f}$ & 304.23717 & -70.750281 & $0.06\pm0.01$ & $0.16\pm0.02$ & $1.78\pm0.09$ & $ 5.45\pm0.28$ & $ 9.3\pm0.5$ & $15.0\pm0.8$ & $22.3\pm1.1$ & $ 33\pm 6$
\enddata
\tablenotetext{a}{Center $^{\rm b}$Tidal Dwarf Galaxy (TDG) candidate $^{\rm c}$Northeastern Arm $^{\rm d}$Body $^{\rm e}$Southwestern Arm $^{\rm f}$Companion}	
\tablenotetext{$\star$}{All fluxes are in mJy and are not yet corrected by foreground Galactic extinction.} 
\tablenotetext{$\star$}{Upper limits and all uncertainties are 1-$\sigma$ uncertainties}
\end{deluxetable*}
%\clearpage
%\end{landscape}
%===================================================================

\setcounter{table}{1}
%========================= Table 2 (cont.) =========================
%\clearpage
%\begin{landscape}
\begin{deluxetable*}{cccccccccccccccccc}
\tabletypesize{\scriptsize}
\tablecaption{(cont.) Spectral Energy Distributions - Part 2}
\tablewidth{0pt}
\tablehead{
Reg& H          & K$_{\rm S}$& W1             & IRAC1          & IRAC2          & W2             & IRAC3        & IRAC4         & W3             & W4 \\
   & (1.7\,\um) & (2.2\,\um) & (3.5\,\um)     & (3.6\,\um)     & (4.5\,\um)     & (4.6\,\um)     & (5.6\,\um)   & (8.0\,\um)    & (12\,\um)      & (22\,\um) }
\startdata
\\ 
 0 & $145\pm15$ & $120\pm13$ & $59.20\pm1.42$ & $59.56\pm2.98$ & $36.01\pm1.80$ & $31.89\pm0.89$ & $30.1\pm1.5$ & $30.0\pm1.5 $ & $18.15\pm0.82$ & $18.20\pm1.04$\\ \\ \hline \\
 1 & $    <  5$ & $    <  6$ & $ 0.15\pm0.03$ &    \nodata     & $      < 0.06$ & $ 0.05\pm0.01$ &   \nodata    & $ 1.1\pm0.1 $ & $ 0.26\pm0.01$ & $ 0.37\pm0.04$\\ \\ \hline \\
 2 & $    <  5$ & $    <  6$ & $ 0.25\pm0.03$ & $ 0.05\pm0.04$ & $ 0.11\pm0.06$ & $ 0.09\pm0.01$ & $     < 0.4$ & $     < 0.03$ & $ 0.19\pm0.01$ & $      < 0.06$\\
 3 & $    <  5$ & $    <  6$ & $ 0.69\pm0.03$ & $ 0.62\pm0.05$ & $ 0.36\pm0.06$ & $ 0.39\pm0.02$ & $     < 0.4$ & $ 1.6\pm0.1 $ & $ 1.31\pm0.06$ & $ 2.50\pm0.15$\\
 4 & $    <  5$ & $    <  6$ & $ 0.73\pm0.03$ & $ 0.68\pm0.05$ & $ 0.46\pm0.06$ & $ 0.38\pm0.02$ & $     < 0.4$ & $ 1.0\pm0.1 $ & $ 0.97\pm0.04$ & $ 1.33\pm0.08$\\
 5 & $    <  5$ & $    <  6$ & $ 1.23\pm0.04$ & $ 1.19\pm0.07$ & $ 0.75\pm0.07$ & $ 0.64\pm0.02$ & $     < 0.4$ & $ 1.6\pm0.1 $ & $ 1.34\pm0.06$ & $ 2.02\pm0.12$\\
 6 & $    <  5$ & $    <  6$ & $ 2.48\pm0.07$ & $ 2.58\pm0.14$ & $ 1.62\pm0.10$ & $ 1.47\pm0.04$ & $ 2.2\pm0.4$ & $ 7.8\pm0.4 $ & $ 4.64\pm0.21$ & $ 6.49\pm0.37$\\
 7 & $ 14\pm 6$ & $ 13\pm 6$ & $ 8.18\pm0.20$ & $ 8.25\pm0.41$ & $ 5.37\pm0.27$ & $ 4.92\pm0.14$ & $12.4\pm0.7$ & $34.8\pm1.7 $ & $20.39\pm0.92$ & $34.39\pm1.96$\\ \\ \hline \\
 8 & $ 42\pm 7$ & $ 35\pm 7$ & $19.08\pm0.46$ & $19.28\pm0.97$ & $11.92\pm0.60$ & $10.56\pm0.30$ & $14.5\pm0.8$ & $28.8\pm1.4 $ & $17.39\pm0.78$ & $19.77\pm1.13$\\
 9 & $ 34\pm 6$ & $ 30\pm 6$ & $16.94\pm0.41$ & $16.31\pm0.82$ & $10.43\pm0.52$ & $ 9.45\pm0.26$ & $12.4\pm0.7$ & $23.1\pm1.2 $ & $14.08\pm0.63$ & $16.17\pm0.92$\\
10 & $ 45\pm 7$ & $ 39\pm 7$ & $20.10\pm0.48$ & $19.69\pm0.99$ & $12.27\pm0.62$ & $11.51\pm0.32$ & $14.2\pm0.8$ & $23.5\pm1.2 $ & $13.87\pm0.62$ & $18.43\pm1.05$\\ \\ \hline \\
11 & $ 26\pm 6$ & $ 23\pm 6$ & $13.95\pm0.34$ & $13.77\pm0.69$ & $ 8.96\pm0.45$ & $ 8.74\pm0.24$ & $18.7\pm1.0$ & $46.0\pm2.3 $ & $25.38\pm1.14$ & $42.64\pm2.43$\\
12 & $ 11\pm 5$ & $ 10\pm 6$ & $ 6.63\pm0.16$ & $ 5.94\pm0.30$ & $ 3.93\pm0.20$ & $ 4.38\pm0.12$ & $ 8.3\pm0.5$ & $14.6\pm0.7 $ & $ 8.67\pm0.39$ & $15.27\pm0.87$\\
13 & $    <  5$ & $    <  6$ & $ 2.80\pm0.07$ & $ 2.12\pm0.11$ & $ 1.51\pm0.09$ & $ 1.72\pm0.05$ & $ 3.2\pm0.4$ & $ 5.2\pm0.3 $ & $ 3.29\pm0.15$ & $ 4.42\pm0.25$\\
14 & $    <  5$ & $    <  6$ & $ 0.87\pm0.03$ & $ 0.60\pm0.05$ & $ 0.54\pm0.06$ & $ 0.53\pm0.02$ & $ 0.4\pm0.4$ & $ 0.8\pm0.1 $ & $ 0.73\pm0.03$ & $ 1.55\pm0.09$\\
15 & $    <  5$ & $    <  6$ & $ 0.63\pm0.03$ & $ 0.42\pm0.05$ & $ 0.34\pm0.06$ & $ 0.46\pm0.02$ & $     < 0.4$ & $ 1.2\pm0.1 $ & $ 0.70\pm0.03$ & $ 0.48\pm0.04$\\ \\ \hline \\
16 & $ 39\pm 7$ & $ 33\pm 6$ & $16.58\pm0.40$ & $16.92\pm0.85$ & $11.33\pm0.57$ & $ 9.92\pm0.28$ & $13.6\pm0.8$ & $27.6\pm1.4 $ & $17.21\pm0.77$ & $25.26\pm1.44$\\
\enddata
\tablenotetext{$\star$}{All fluxes are in mJy and uncorrected for foreground Galactic extinction.} 
\tablenotetext{$\star$}{Upper limits and all uncertainties are 1-$\sigma$ uncertainties}
\end{deluxetable*}
%\clearpage
%\end{landscape}
%===================================================================

%%%%%%%%%%%%%%%%%%%%%%%%%%%%%%%%%%%%%%%%%%%%%%%%%%%%%%%%%%%%%%%%%%%
\section{GENERAL SPECTRAL PROPERTIES AND MORPHOLOGY} \label{sec:SEDs}
%%%%%%%%%%%%%%%%%%%%%%%%%%%%%%%%%%%%%%%%%%%%%%%%%%%%%%%%%%%%%%%%%%%
\subsection{Spectral Energy Distributions} \label{sec:SEDss}
\begin{figure}[b]
%\begin{center}
\hspace{-0.35in}
\includegraphics[width=.55\textwidth]{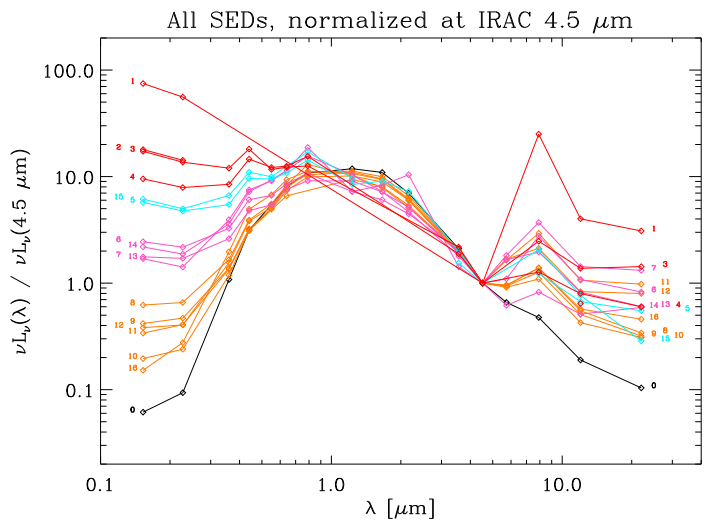}
\caption{SEDs of all regions normalized by their \spitzer\ 4.5\,\um\ intensity. They are illustrated in the color-code defined in \S~\ref{sec:SEDss}. Region numbers are shown accompanying the beginning and end-points of the SEDs.}
\label{fig:SEDs_norm}
%\end{center}
\end{figure}

%%%%%%%%%%%%%%%%%%%%%%%%%%%%%%%%%%%%%%%%%%%%%%%%%%%%%%%%%%%%%%%%%%%
The SEDs of the 17 regions are shown in Figure~\ref{fig:SEDs_norm}. They have been all normalized to the same intensity at 4.5~\um\ in order to emphasize the different relative contributions of old and young stellar populations to the SEDs in different regions of the galaxy.

%============================ Figure 03 ============================
%===================================================================

Based on the figure, we have sorted the SEDs of the 17 regions into 5 groups according to their FUV/4.5~\um\ flux ratios . The groups are color-coded in Figure~\ref{fig:DSS_FUV_regions}, and show a trend of redder colors at smaller distances from the nucleus, across both arms of the galaxy. The trend suggests a symmetric variation in the ages of stellar populations, with the exception of the outermost regions of the East arm. The figure shows that the normalized FUV flux of the the TDG candidate is 3 orders of magnitudes brighter than that of the central region of the galaxy (Region 0). Regions 1--4 have much bluer UV-4.5~\um\ colors than all other regions. 

The bluest region, designated Region 1, resembles a tidal dwarf galaxy (TDG). TDGs are gravitationally bound recycled objects formed out of material tidally pulled from interacting parent galaxies in the intergalactic medium \citep{DM1998}. A rotation curve is necessary to show the object is bound. In order for it to be a TDG formed from the pre-enriched gas of the main spiral, the galaxy has to have considerably high metallicity \citep[$12+\log({\rm O}/{\rm H})>8.3$,][]{DM1998,Boquien2009}. We do not have spectroscopic data to test either of these conditions, or even if in fact Region 1 is detached from the extended disk of the Condor. Therefore, we refer to Region 1 as a TDG-candidate. Later in the paper we will derive its stellar mass and SFR.

We expect the SEDs of all regions to arise from a composite stellar population, consisting of an old pre-collision component and a younger one formed by the interaction. The latter will give rise to most of the UV flux. Consequently, based solely on the observed FUV-NUV colors (see slopes in Figure~\ref{fig:SEDs_norm}), the companion (Region 16) seems to have the oldest stellar population with an SED that is similar to that of regions 0 and 10, in the central regions of the Condor.

%%%%%%%%%%%%%%%%%%%%%%%%%%%%%%%%%%%%%%%%%%%%%%%%%%%%%%%%%%%%%%%%%%%
\subsection{The Bar Region}\label{sec:bar}
%%%%%%%%%%%%%%%%%%%%%%%%%%%%%%%%%%%%%%%%%%%%%%%%%%%%%%%%%%%%%%%%%%%
The Condor has a particularly long bar. To determine its size, we performed a 2-D bulge/bar/disk decomposition of the Condor using the \spitzer\ 3.6\,\um\ image and the {\sc budda} software package (\citealt{deSouza2004}; \citealt{Gadotti2008}), which indicates that the disk profile has an outer break, from where the profile is shallower than inwards, i.e. the disk is a type III disk (see \citealt{Erwin2008}; \citealt{MunozMateos2013}). The decomposition gives a bar semi-major axis of $\sim10$~kpc, and a disk scale length of $\sim 7$~kpc. Both the bar semi-major axis and disk scale length are quite large, befitting a giant galaxy \cite[their median values in the local universe are 4.5\,kpc and 2.8\,kpc, respectively, as in][]{Gadotti2009}. The ratio of the bar semi-major axis to the disk scale-length found for non-interacting barred galaxies is $1.5 \pm 0.2$ \citep[][see Figure 1 in his paper]{Gadotti2011}. For the Condor this ratio is 1.4, a value typically found in the local universe. This could be an indication that the length of the bar has not been abruptly affected by the interaction with the companion. Alternatively, the interaction could have increased the disk scale-length accordingly, keeping to some extent constant the ratio between bar length and disk scale-length.

\cite{Phillips1996} shows evidence that in some barred galaxies star formation occurs throughout the bar, whereas in other barred galaxies star formation is confined to the bar center and ends. The Condor falls in the second category, since its \ha\ \citep{Mihos1993}, UV, mid-IR  maps suggest the star formation is confined to the ends of the bar and spiral arms. Comparing the SEDs of the nucleus and regions of the bar on 5\arcsec\ scale (1.5~kpc), we found them to be almost identical and therefore equally old. This suggests an evolved bar, and an old pre-collision stellar population in both the bar and the nucleus. There is therefore no spectral signature of a young population in the bar. We also find no signs of a box-peanut structure near the central region. The absence of such morphological feature would indicate a young bar, i.e. one with a dynamical age less than about 2 Gyrs. However, the lack of such a structure could be a projection effect and should not be taken as conclusive evidence for the age of the bar. Altogether, the evidence points to a bar that formed a few billion years before the interaction, as will be derived in a later section. It also suggests that the bar was not fed with gas at the interaction.

The fact that star formation activity is not concentrated along the bar and central regions of the galaxy presents itself as a major challenge to the current simulations describing the state of the galaxy. \cite{Mihos1993} pointed out an improvement to the model in which the Condor had a uniform gas disk. If prior to the interaction with the companion the \hi\ gas was distributed at larger radii, with a central region devoid of \hi\ gas instead, star formation would be strongest along the spiral arms. This might be the case here, and future simulations should explore this possibility in more details.

A bar older than the interaction with the companion implies it was not formed by this interaction. The companion must have played a smaller role in shaping the Condor than previously thought. The more distant elliptical companion NGC~6876 could have shaped the Condor over a longer period of time, possibly distorting its disk and contributing to the formation of the bar instability much before the interaction with the companion started, in the last 130~Myr.

%%%%%%%%%%%%%%%%%%%%%%%%%%%%%%%%%%%%%%%%%%%%%%%%%%%%%%%%%%%%%%%%%%%
\section{DETERMINING THE STAR FORMATION HISTORIES} \label{sec:SFH}
%%%%%%%%%%%%%%%%%%%%%%%%%%%%%%%%%%%%%%%%%%%%%%%%%%%%%%%%%%%%%%%%%%%

%%%%%%%%%%%%%%%%%%%%%%%%%%%%%%%%%%%%%%%%%%%%%%%%%%%%%%%%%%%%%%%%%%%
\subsection{SED Fitting} \label{sec:SED_fittings}
%%%%%%%%%%%%%%%%%%%%%%%%%%%%%%%%%%%%%%%%%%%%%%%%%%%%%%%%%%%%%%%%%%%

Modeling the star formation rate as function of time, i.e. the SFH, of a complex stellar population is the key to obtain a reliable stellar mass and current SFR. From numerical simulations, interactions episodes between galaxies are usually followed by an enhancement in star formation that decays roughly exponentially, as seen in \cite{Mihos1993}, with decay constant varying from case to case.

In the last decades, with the large availability of multi-wavelength data and the improvements in evolutionary population synthesis codes, modeling the SFH became necessary in order to describe the SEDs of all types of galaxies (see reviews by \citealt{Walcher2011}, \citealt{Conroy2013}, and references therein). When modeling interacting systems, previous works usually assume a superposition of one or two exponential decays and/or single instantaneous bursts for whole galaxies, sub-galactic regions, or clumps \citep{Smith2014,MC2012,Boquien2009,Boquien2010,Boquien2011}.

Since the Condor is a spiral galaxy we model the SFH as a single component represented by an exponential function:

%---------------
\begin{equation}
  \psi_1(t) =  
    \left\{ 
      \begin{array}{c l l}
        0                                                &  & {\rm for }\  0  \le t  <  t_i \\
        \left(M_{\infty}/{\tau}\right)~e^{-(t-t_i)/\tau} &  & {\rm for }\ t_i \le t \le t_0 {\ \rm ,}
      \end{array}
    \right.
\end{equation}
%---------------
where $t$ is cosmic time, $t_i$ is the age of the universe at the onset of star formation, $t_0$ is the current epoch, so that $t_{\rm age} = t_0 - t_i$ is the age of the region, $\tau$ is the decay time, and $M_{\infty}$ is the total gas mass consumed by star formation as $t_{\rm age}$ approaches infinity. 

We used the population synthesis code P\'EGASE.2 \citep{FRM1999} to generate the SEDs of the stellar populations, for a fixed solar metallicity ($Z=0.02$) and a Kroupa initial mass function (IMF), given by \citep{Kroupa2001}:
%---------------
\begin{equation}
  \Phi(m) = \frac{{\rm d}N}{{\rm d}m} \propto
    \left\{ 
      \begin{array}{l l}
        m^{-1.3}{\rm ,} & {\rm for }\ 0.1 \le m/\msun \le 0.5 \\
        m^{-2.3}{\rm ,} & {\rm for }\ 0.5 \le m/\msun \le 100 {\ \rm .}
      \end{array}
    \right.
  \label{eq:KroupaIMF}
\end{equation}
%---------------
The specific luminosity is then
\begin{equation}
  L_{\nu}(\lambda) = M_{\infty}~{\widetilde L}_{\nu}(\lambda,\tau,t_{\rm age})
                     ~ e^{-\tau_{\rm int}(\lambda)} {\quad\rm ,}
  \label{eq:model}
\end{equation}
where ${\widetilde L}_{\nu}{(\lambda,\tau,t_{\rm age})}$ is the SED template obtained for a value of $M_{\infty} = 1\,\msun$.  The optical depth $\tau_{\rm int}(\lambda)$ is the intrinsic extinction in the source, taken to be represented by the \cite{Calzetti1997} extinction law. The model has thus 4 free parameters: $M_{\infty}$, $\tau$, $t_{\rm age}$, and $\tau_{\rm int}({\rm V})$, the intrinsic extinction in the V-band, hereafter expressed in terms of $A_{\rm V}$.

%============================= Table 3 =============================
\begin{deluxetable*}{cccccccccc}[p]
\tabletypesize{\scriptsize}
\tablecaption{SED fitting Results \label{tab:results_4}}
\tablewidth{0pt}
\tablehead
{
 \multicolumn{1}{c}{} &        &                 &        &               &        &                     &                  &                 \\
 [\dimexpr-\normalbaselineskip-\arrayrulewidth]% Correct for mis-alignment
        & \vline & \multicolumn{4}{c}{Model Parameters}                   & \vline & \multicolumn{3}{c}{Derived Quantities}                   \\
 Region & \vline & A$_{\rm V}$ & $M_{\infty}$    & $\tau$ & $t_{\rm age}$ & \vline & $\left<\psi\right>$ & $M_{\star}$      & $L_{\rm bol}$   \\
 $[\#]$ & \vline & [mag]       & [$10^9\,\msun$] & [Gyr]  & [Gyr]         & \vline & [$10^{-2}\msunyr$]  &  [$10^9\,\msun$] & [$10^9\,\msun$]
}
\startdata
 0 & \vline & $0.27^{+0.13}_{-0.11}$ & $118.53^{+35.52}_{-22.74}$ & $0.53^{+0.22}_{-0.13}$ & $4.38^{+1.92}_{-1.16}$ & \vline & $  6.6^{+  2.7}_{- 1.5}$ & $60.83^{+14.89}_{- 9.90}$ & $66.21^{+3.51}_{-2.78}$\\
   & \vline & & & & & \vline & & & \\ 
 1 & \vline & $0.53^{+0.67}_{-0.31}$ & $  0.01^{+ 0.03}_{- 0.00}$ & $0.00^{+0.07}_{-0.00}$ & $0.02^{+0.06}_{-0.01}$ & \vline & $  8.1^{+ 30.0}_{- 4.8}$ & $ 0.01^{+ 0.02}_{- 0.00}$ & $ 0.68^{+0.86}_{-0.15}$\\
   & \vline & & & & & \vline & & & \\ 
 2 & \vline & $0.45^{+0.38}_{-0.07}$ & $  0.01^{+ 0.01}_{- 0.00}$ & $0.00^{+0.03}_{-0.00}$ & $0.02^{+0.02}_{-0.01}$ & \vline & $ 12.3^{+  5.6}_{- 4.2}$ & $ 0.01^{+ 0.00}_{- 0.00}$ & $ 1.02^{+1.24}_{-0.09}$\\
   & \vline & & & & & \vline & & & \\ 
 3 & \vline & $0.63^{+0.16}_{-0.25}$ & $  0.12^{+ 0.06}_{- 0.01}$ & $0.05^{+0.12}_{-0.03}$ & $0.10^{+0.04}_{-0.01}$ & \vline & $104.2^{+ 18.0}_{-41.5}$ & $ 0.08^{+ 0.01}_{- 0.01}$ & $ 4.99^{+1.63}_{-2.01}$\\
   & \vline & & & & & \vline & & & \\ 
 4 & \vline & $0.26^{+0.11}_{-0.10}$ & $  0.41^{+ 0.12}_{- 0.10}$ & $0.29^{+0.34}_{-0.15}$ & $0.67^{+0.42}_{-0.26}$ & \vline & $ 16.7^{+  5.1}_{- 3.7}$ & $ 0.24^{+ 0.05}_{- 0.04}$ & $ 2.43^{+0.42}_{-0.29}$\\
   & \vline & & & & & \vline & & & \\ 
 5 & \vline & $0.21^{+0.13}_{-0.15}$ & $  1.37^{+ 0.27}_{- 0.17}$ & $1.21^{+0.81}_{-0.33}$ & $2.50^{+0.88}_{-0.48}$ & \vline & $ 15.0^{+  5.4}_{- 4.0}$ & $ 0.69^{+ 0.11}_{- 0.09}$ & $ 2.88^{+0.52}_{-0.40}$\\
   & \vline & & & & & \vline & & & \\ 
 6 & \vline & $0.66^{+0.10}_{-0.13}$ & $  2.48^{+ 0.54}_{- 0.32}$ & $0.86^{+0.49}_{-0.22}$ & $1.84^{+0.71}_{-0.39}$ & \vline & $ 35.9^{+  9.1}_{- 8.7}$ & $ 1.31^{+ 0.24}_{- 0.15}$ & $ 6.64^{+0.85}_{-0.88}$\\
   & \vline & & & & & \vline & & & \\ 
 7 & \vline & $0.88^{+0.10}_{-0.11}$ & $  9.66^{+ 2.27}_{- 1.68}$ & $1.52^{+1.04}_{-0.51}$ & $2.29^{+0.79}_{-0.59}$ & \vline & $145.7^{+ 30.5}_{-27.9}$ & $ 4.43^{+ 0.63}_{- 0.57}$ & $24.26^{+2.88}_{-2.73}$\\
   & \vline & & & & & \vline & & & \\ 
 8 & \vline & $0.56^{+0.08}_{-0.12}$ & $ 25.14^{+ 2.72}_{- 2.75}$ & $0.55^{+0.11}_{-0.08}$ & $2.48^{+0.54}_{-0.41}$ & \vline & $ 54.0^{+ 12.5}_{-13.2}$ & $13.90^{+ 1.18}_{- 1.25}$ & $29.07^{+1.89}_{-2.12}$\\
   & \vline & & & & & \vline & & & \\ 
 9 & \vline & $0.69^{+0.07}_{-0.12}$ & $ 22.51^{+ 2.32}_{- 2.24}$ & $0.53^{+0.08}_{-0.07}$ & $2.49^{+0.48}_{-0.35}$ & \vline & $ 43.5^{+  7.3}_{-10.4}$ & $12.45^{+ 1.00}_{- 1.04}$ & $25.36^{+1.20}_{-1.84}$\\
   & \vline & & & & & \vline & & & \\ 
10 & \vline & $0.54^{+0.16}_{-0.10}$ & $ 30.56^{+ 2.14}_{- 4.62}$ & $0.47^{+0.07}_{-0.11}$ & $2.86^{+0.45}_{-0.73}$ & \vline & $ 15.9^{+  7.4}_{- 2.5}$ & $16.59^{+ 0.86}_{- 2.00}$ & $26.00^{+2.25}_{-1.24}$\\
   & \vline & & & & & \vline & & & \\ 
11 & \vline & $1.01^{+0.25}_{-0.07}$ & $ 12.32^{+ 1.62}_{- 0.23}$ & $0.28^{+0.04}_{-0.03}$ & $1.30^{+0.19}_{-0.25}$ & \vline & $ 54.4^{+ 38.0}_{-10.9}$ & $ 7.39^{+ 0.82}_{- 0.13}$ & $24.86^{+5.53}_{-1.24}$\\
   & \vline & & & & & \vline & & & \\ 
12 & \vline & $0.78^{+0.06}_{-0.09}$ & $  7.10^{+ 0.95}_{- 0.66}$ & $0.31^{+0.06}_{-0.04}$ & $1.60^{+0.35}_{-0.17}$ & \vline & $ 16.4^{+  3.1}_{- 3.1}$ & $ 4.14^{+ 0.44}_{- 0.33}$ & $10.89^{+0.36}_{-0.57}$\\
   & \vline & & & & & \vline & & & \\ 
13 & \vline & $0.22^{+0.08}_{-0.15}$ & $  2.88^{+ 0.51}_{- 0.35}$ & $0.35^{+0.08}_{-0.05}$ & $1.68^{+0.49}_{-0.26}$ & \vline & $  8.0^{+  2.2}_{- 2.1}$ & $ 1.67^{+ 0.24}_{- 0.17}$ & $ 4.44^{+0.26}_{-0.39}$\\
   & \vline & & & & & \vline & & & \\ 
14 & \vline & $0.24^{+0.26}_{-0.10}$ & $  0.73^{+ 0.21}_{- 0.03}$ & $0.31^{+0.13}_{-0.06}$ & $1.32^{+0.54}_{-0.28}$ & \vline & $  3.8^{+  2.9}_{- 0.8}$ & $ 0.44^{+ 0.10}_{- 0.01}$ & $ 1.52^{+0.38}_{-0.09}$\\
   & \vline & & & & & \vline & & & \\ 
15 & \vline & $0.46^{+0.25}_{-0.04}$ & $  0.12^{+ 0.10}_{- 0.02}$ & $0.00^{+0.08}_{-0.00}$ & $0.11^{+0.15}_{-0.02}$ & \vline & $  0.1^{+119.6}_{- 0.1}$ & $ 0.10^{+ 0.05}_{- 0.01}$ & $ 1.54^{+0.91}_{-0.05}$\\
   & \vline & & & & & \vline & & & \\ 
16 & \vline & $0.38^{+0.02}_{-0.07}$ & $ 17.75^{+ 1.97}_{- 1.10}$ & $0.21^{+0.03}_{-0.01}$ & $1.49^{+0.21}_{-0.07}$ & \vline & $  7.8^{+  0.9}_{- 1.5}$ & $10.40^{+ 0.96}_{- 0.58}$ & $24.11^{+0.33}_{-0.62}$\\
\enddata
\end{deluxetable*}
%===================================================================

Specific luminosities were convolved with the transmission curve of the filter with the effective wavelength $\lambda_k$ of each instrument in order to derive the broad-band SED, $L_\nu(\lambda_k)$. They were corrected for foreground Galactic UV and optical extinction using the Milky Way extinction curve for a value of $A_V^{\rm Gal}=0.127$ \citep{Schlafly2011, Schlegel1998}. At FUV and NUV wavelengths we used an extinction correction given by: $A_{\rm FUV}^{\rm Gal} = 8.29\times E(B-V)$ and $A_{\rm NUV}^{\rm Gal} = 8.18 \times E(B-V)$, with $E(B-V) = 0.041$ \citep{Seibert2005}. 

We have generated a four-dimensional grid of the parameter space $\left\{A_{\rm V},\, M_{\infty},\, \tau,\, t_{\rm age}\right\}$. The grid is linearly spaced in the intrinsic value of $A_{\rm V}$ and logarithmically spaced in each the other three parameters, with 100 steps in each parameter (adding up to 100 million models), spanning the following ranges of parameter space: 
%---------------
\begin{equation}
  \begin{array}{c}
    0.01    \le A_{\rm V}/{\rm mag} \le 2.0 \quad, \\
    10^{7}  \le M_{\infty}/\msun \le 10^{12} \quad, \\ 
    0.01    \le \tau/{\rm Gyr} \le 10 \quad, \\
    0.01    \le t_{\rm age}/{\rm Gyr} \le 13.8 \quad.\\
  \end{array}
  \label{eq:ranges}
\end{equation}

\begin{figure*}[h]
\begin{center}
  \includegraphics[width=1.\textwidth]{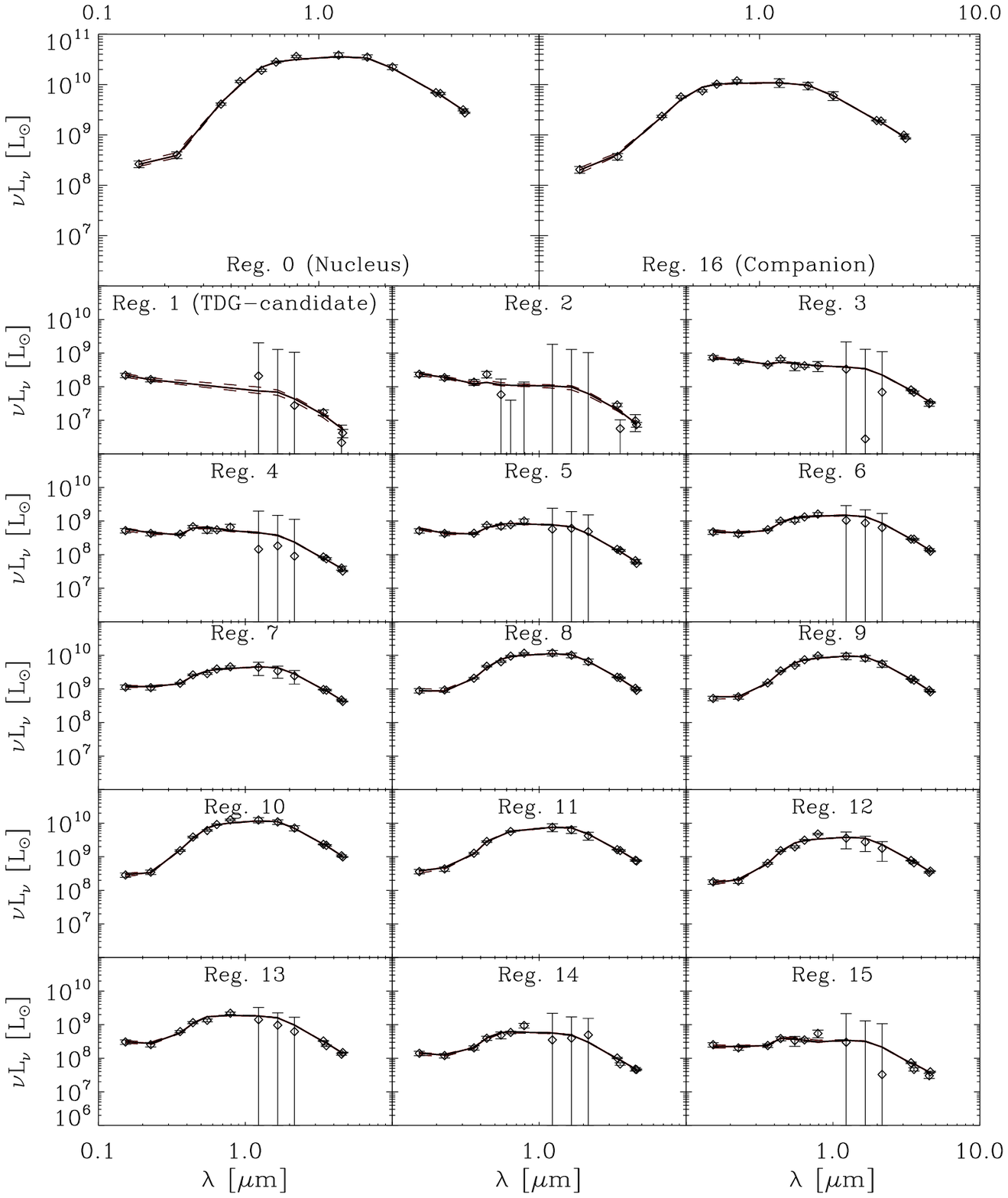}
    \caption{\footnotesize Observed SEDs of all 17 regions compared to the models. Open black diamonds are the measurements and their uncertainties. The lowest-$\chi^2$ models are shown as the solid curves. Upper and lower 1-$\sigma$ bounds of the solid curves are shown as dashed curves.}
  \label{fig:All_SEDs_4par}
\end{center}
\end{figure*}

We fit the region's SED by first selecting the grid parameters giving the lowest value of:
\begin{equation}
  \chi^2 = \sum_k{\frac{1}{\sigma_k^2}\left[L_\nu(\lambda_k) - L_\nu^{\rm obs}(\lambda_k)\right]^2}{\ \rm.}
\end{equation}
The process was then repeated using a finer grid encompassing a 1-$\sigma$ range of the best fitting parameters of the coarse grid.
The parameters and derived quantities of the best fitting model SEDs are presented in Table \ref{tab:results_4}.
The uncertainties in the parameters were obtained by projecting the volume contained within $\chi^2 \leq \chi^2_{\rm min}+1$ onto the five axes of the parameter space. Figure \ref{fig:All_SEDs_4par} shows the results of the SED fittings for all 17 regions. The solid curves are the best-fit models and dashed curves are their 1-$\sigma$ bounds.

%============================ Figure 04 ============================
%===================================================================

We also performed SED fittings adding another component to the SFHs. This component was chosen as a constant SFR for the last 100~Myr as an attempt to separately capture any enhancements in the star formation rate caused by the interaction. The addition of this one-parameter component, however, fails an F-test \citep{Bevington} at the 99\% confidence level for all regions. Therefore, the added starburst component does not significantly improve the fits. In fact, the exponential decay already captures both the old and young stellar populations for each region of the galaxy. In the following, we base all our discussions on the simple 4-parameter model described in \S~\ref{sec:SED_fittings}. 

%============================================
\subsection{Large-Scale Trends}\label{sec:trends}
\begin{figure*}
\begin{center}
  \includegraphics[width=1.0\textwidth]{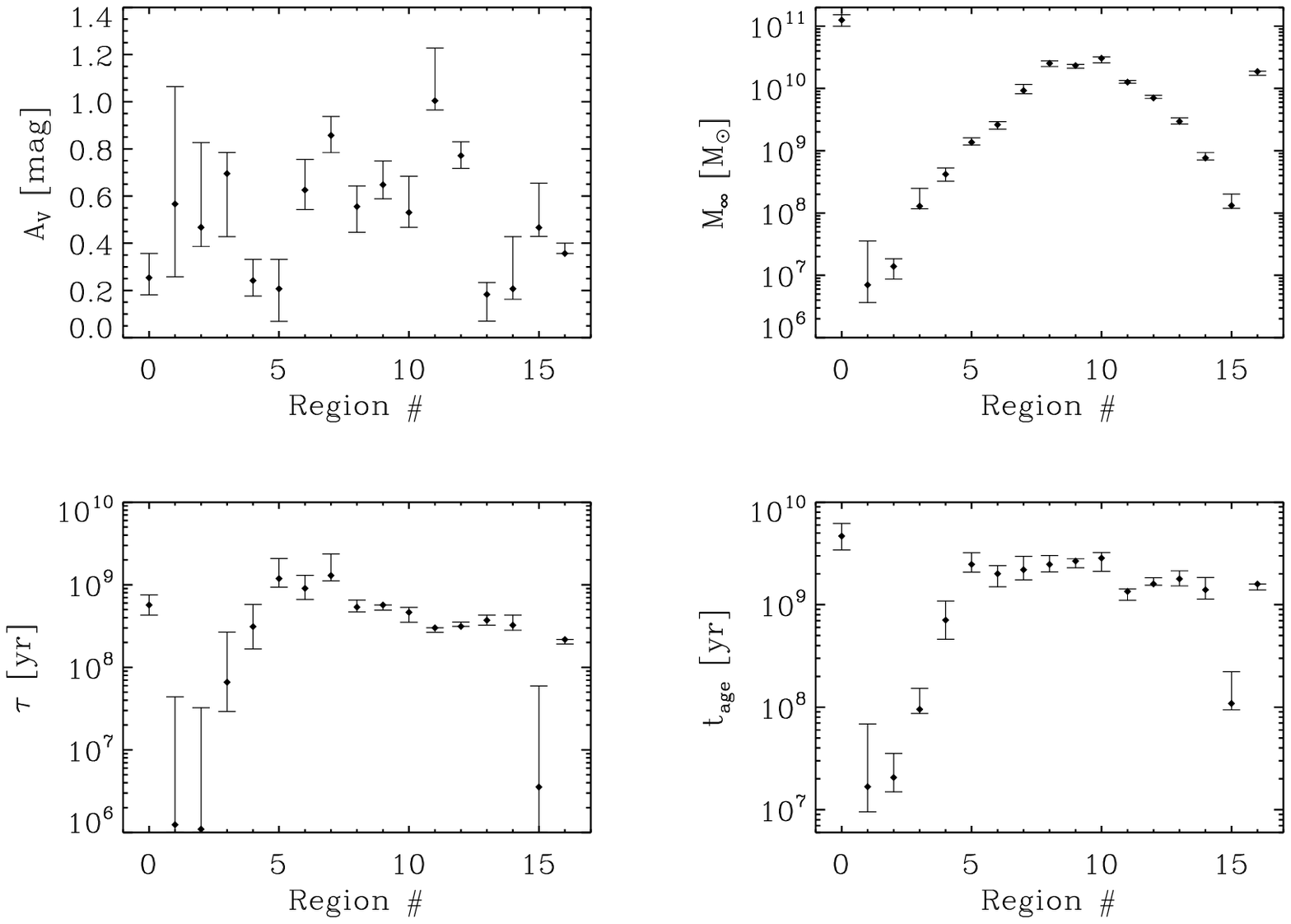}
  \caption{\footnotesize Parameters derived from the models. Values in Table~\ref{tab:results_4}. Discussion in Section~\ref{sec:trends}.}
  \label{fig:parameters_45}
  \end{center}
\end{figure*}

\begin{figure*}
\begin{center}
  \includegraphics[width=1.\textwidth]{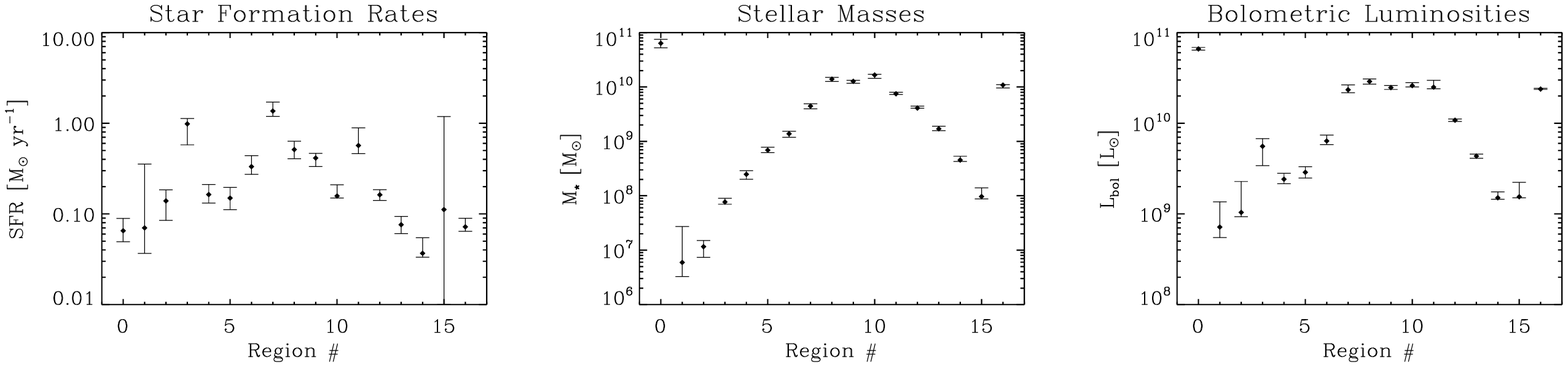}
  \caption{\footnotesize Average SFR of the last 100~Myr, stellar masses, and bolometric luminosity of all the regions, as derived from the models. Values in Table~\ref{tab:results_4}. Discussion in Section~\ref{sec:trends}.}
  \label{fig:derived_45}
\end{center}
\end{figure*}

%============================================
Inspection of Figures \ref{fig:parameters_45} and \ref{fig:derived_45} shows interesting large scale trends of the properties of the 17 regions investigated here. Regions 3 and 15 seem to be as old as the interaction between the Condor and the companion, with $t_{\rm age}$ of the order of 100~Myr. Regions 1 and 2 are younger than the interaction, with $t_{\rm age} \sim 20$~Myr, and seem to have been fully assembled after the closest approach of the companion. Region 4 has an age around 700~Myr and all other regions are older than a gigayear. The central region of the Condor (Region 0) is the oldest region, with $t_{\rm age} \sim 4.4$~Gyr.

%============================ Figure 05 ============================
%===================================================================

%============================ Figure 06 ============================
%===================================================================

As shown in the previous section, the SFH of all regions can be characterized by a single declining SFR. If a region had a constant SFH, it would be manifested by $\tau/t_{\rm age}  >> 1$. However, this is not the case for any of the observed regions, all of them have $\tau/t_{\rm age} < 1$. An ``instantaneous" burst will have $\tau/t_{\rm age} << 1$. Regions 1, 2, and 15, at the outermost parts of the galaxy, are the closest to be characterized as such a burst, with a ratios lower than $0.1$.

The 100~Myr-averaged SFR $\left<\psi\right>$ summed over all Regions in the Condor (0-15) is equal to $5.4~\msunyr$ and equal to $0.08~\msunyr$ in the companion. We derive a total stellar mass of $1.2^{+0.2}_{-0.1}\times10^{11}~\msun$ for Regions 0-15 and $1.0^{+0.1}_{-0.1}\times10^{10}~\msun$ for the companion. With this mean SFR over the past 100~Myr, all regions converted around $5.5\times10^{8}~\msun$ of gas into stars. This gas mass is negligible compared to the $2.15\times10^{10}~\msun$ \hi\ reservoir of the whole Condor/companion system \citep{HB1997}. This places both the Condor and the companion on the main sequence of star formation {\rm \citep{Wijesinghe2011,Elbaz2011,Noeske2007}}, with specific star formation rates (sSFRs) of $4.5\times10^{-11}~{\rm yr}^{-1}$ and $8\times10^{-12}~{\rm yr}^{-1}$, respectively, which corresponds to stellar mass doubling times of $2.2\times10^{10}~{\rm yr}$ and $1.3\times10^{11}~{\rm yr}$. 

The SFR derived here is considerably lower than the $\sim 27~\msunyr$ derived by \cite{Bastian2005} solely from the U-band. However, we consider our lower SFR, derived from modeling the SED from the FUV to near-IR, more reliable that based on a single photometric band. The SFRs derived here are consistent with the ones derived by \cite{Mihos1993}, who estimated a ${\rm SFR} < 10.9~\msunyr$ for the whole galaxy\footnote{originaly ${\rm SFR}<5.7~\msunyr$, at a distance of 47~Mpc}, based on the \IRAS\ IR colors. They noted the galaxy was cold compared to other interacting galaxies and that possibly a considerable fraction of this IR emission is not associated with the recent SFR (hence the upper limit). In fact, we find this fraction to be roughly equal to 0.5. On the other hand, a compound SFR derived in \cite{Hao2011} including FUV and \wise\ 22~\um\ (or NUV and \wise\ 22~\um) luminosities leads to a SFR of $3.4~\msunyr$ (or $3.3~\msunyr$) for all our regions combined. We discuss the possible physical origins for the discrepancies between our SED-fitting SFRs and the ones combining FUV and mid-IR in the next sub-sections.

The ratio of stellar masses between the Condor and its companion derived here is greater than 12, considerably higher than 5, the value used by previous numerical simulations capable of reproducing the current state of the system \citep{Mihos1993,HK07}. The initial attempt at modeling this system by \cite{Mihos1993} assumed a ratio of 10, which was implied from the B-band luminosities. However, they were unable to produce the thin tidal features inferred from the optical, and switched to a ratio of 5 instead. Their simulation also produced mainly star formation in the central areas of the galaxy, but the current available (\ha, FUV, and 22~\um) observations point to a different star formation morphology, with a very small fraction of the total SFR coming from the central region. 

As previously mentioned in \S\ref{sec:bar}, \cite{Mihos1993} discussed another interaction scenario consisting initially not of a uniform gas disk for the Condor, but of a disk with mass distributed only in its outskirts, devoid of gas within galactocentric distances up to 60\% of the disk radius. Under these initial conditions, star formation would not be confined to the bar and central regions of the galaxy, but mainly taking place in the tidally disrupted spiral arms, which is the scenario supported by the current observations. Considering the star formation morphology of the Condor and the stellar mass ratio between the galaxies, it may be necessary to explore the possibility that the disk of the Condor was already more extended before the passage of the companion and the effects of the interaction were not as strong as previously expected. Furthermore, the role of the elliptical galaxy in the interaction may not be negligible, and may be in part responsible for the current morphology of the Condor. Further numerical simulations including star formation are necessary to address these issues.

Region 0 was found to enclose a quite low SFR of $\sim0.07~\msunyr$. A mass of ${\rm H}_2$ molecular gas, $\sim10^9~\msun$, was derived from a single pointing CO observation with a 43\arcsec\ FWHM beam by \cite{HB1997}. Their beam size is larger than the 32\arcsec\ diameter of Region 0. Since no CO map is available for the galaxy, the spatial distribution of the molecular gas is unclear. Nevertheless, we can estimate the mass of molecular gas present in the central region of the Condor using the \cite{Lada2012} relation between the SFR and the mass of dense, $n{\left({\rm H}_2\right)}\ge10^4~{\rm cm^{-3}}$, molecular gas effectively forming stars, given by
\begin{equation}
  \left(\frac{M_{\rm H_2}}{\msun}\right) = \frac{2.2\times10^{7}}{f_{\rm DG}}~\left(\frac{\rm SFR}{\msunyr}\right){\quad\rm,}
\end{equation}
where $f_{\rm DG}$ is the fraction of molecular gas in this dense phase and $M_{\rm H_2}$ is the total molecular gas mass. If all the $1.1\times10^9$\,\msun\ of molecular gas is indeed located in Region~0, then only a small fraction $f_{\rm DG}=1.4\times10^{-3}$ can be in a dense phase. In this case, the molecular gas would be segregated from the \hi\ gas. On the other extreme, if all the molecular gas in Region~0 is in a dense phase, i.e. $f_{\rm DG}=1$, then only $1.5\times10^6~\msun$ of molecular gas can be present in Region 0 to sustain its SFR. This would be consistent with the deficiency of \hi\ gas in Region~0, as observed by \cite{HK07}. A CO map, with higher angular resolution, would be crucial to point determine the distribution of dense molecular gas in the galaxy.

Our analysis makes clear the Condor galaxy was already very massive ($M_{\star}>10^{11}~\msun$) prior to the encounter with the companion, which disturbed the morphology of the galaxy, but did not strongly affect its stellar or gas masses. The Condor was possibly more extended and was less affected by the interaction with the companion IC~4970 than previously thought. The passage did trigger star formation in the system, as clearly seen by the bursts in Regions 1 and 2, with $t_{\rm age} \sim 20$~Myr and $\tau << t_{\rm age}$, but quite low star formation efficiencies are expected in the central region. Modeling the SEDs on smaller scales may reveal more localized effects of the collision.

Another issue of interest is the nature of Region 1, the bluest and outermost region, which could be a TDG. We have derived estimates the stellar mass of $10^7~\msun$ and SFR around $0.1~\msunyr$. TDGs and TDG-candidates with similar SFRs and stellar masses can be found  in \cite{LW2012,Bournaud2009, SRH2012}, but this is not enough to to confirm it nor rule it out as a TDG. As mentioned previously, rotation curve and metallicity are necessary to confirm it as TDG. We, therefore, still refer to it as a TDG-candidate.

%%%%%%%%%%%%%%%%%%%%%%%%%%%%%%%%%%%%%%%%%%%%%%%%%%%%%%%%%%%%%%%%%%%
\section{RESULTS AND DISCUSSION} \label{sec:results}
%%%%%%%%%%%%%%%%%%%%%%%%%%%%%%%%%%%%%%%%%%%%%%%%%%%%%%%%%%%%%%%%%%%
%============================================
\subsection{The FUV + IR emission as a tracer of the SFR}\label{sec:FUVIR}
%============================================
The FUV emission traces the most massive, intrinsically young stars. The intrinsic FUV emission from stars is then a very good tracer of the current SFR. It is, however, attenuated by dust and therefore the observed FUV emission needs to be corrected for extinction before it is converted into a SFR. The dust present in star forming regions and in the interstellar medium (ISM), reprocesses the UV photospheric emission and reemits part of it in the IR. In an attempt to better trace SFR than from the observed FUV luminosity alone, one can complement it with by a single mid-IR band (e.g. {\it IRAS} 25~\micron\ or \spitzer\ 24~\micron).
The choice of 25~\mic\ emission is motivated by the fact that most of the emission at these wavelengths arises from dust that is heated by ionizing and non-ionizing UV photons associated with star formation activity. If the FUV extinction arises solely from the dust giving rise to the mid-IR emission, then the extinction corrected FUV luminosity can be written as:
\begin{equation}
  \nu L_\nu ({\rm FUV})_{\rm corr} = 
  \nu L_\nu ({\rm FUV})_{\rm obs} + a_{\rm corr}\times\nu L_\nu ({\rm 25~\micron})_{\rm obs} {\ \rm,}\label{eq:FUV_corr}
\end{equation}
where \acorr\ is an empirically determined coefficient that related the dust emissivity at 25 micron to a FUV absorptivity. For this work, we use \wise\ 22~\micron\ as our mid-IR band, since no \spitzer\ MIPS data are available for this galaxy and {\it IRAS} data do not have enough spatial resolution to distinguish between our regions. \cite{Lee2013} shows that the \wise\ 22~\um\ and \spitzer\ 24~\um\ data are nearly equivalent for obtaining the unobscured SFRs to complement the obscured \ha\ SFR.
For a large sample of regular galaxies, \cite{Hao2011} found \acorr=$3.89\pm0.15$, using {\it IRAS} 25~\um\ as the mid-IR band.

The relation between the SFR and the intrinsic FUV luminosity is given by
\begin{equation} {\label{eq:SFR_FUV}}
  \left(\frac{\rm SFR}{\msunyr}\right)
   = 1.62\times10^{-10}\left(\frac{\nu L_\nu ({\rm FUV})_{\rm intr}}{{\rm L}_\sun}\right) {\ ,}
\label{eq:FUV_SFR}
\end{equation}
where we used the PEGASE.2 population synthesis code with a Kroupa IMF (Eq. \ref{eq:KroupaIMF}) for a solar metallicity (Z=0.02) and a constant SFR for a duration of 100~Myr. Murphy et al. (2011) derive a coefficient of $1.70\times10^{-10}$ for this relation, using Starburst99 (Leitherer et al. 1999). The difference between the two is negligible compared to uncertainties in the SFRs and FUV luminosities. 

\begin{figure}[h]
%\begin{center}
  \hspace{-0.4in}
  \includegraphics[width=.55\textwidth]{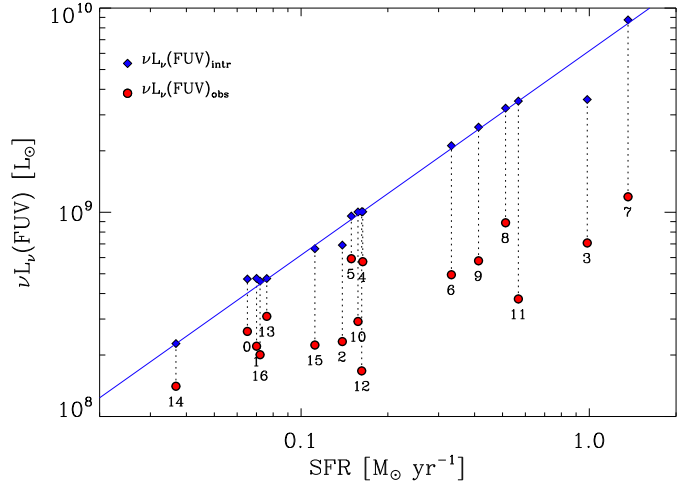}
  \caption{\footnotesize Relation between the model-derived SFR (averaged over the past 100 Myr) and various measures of the FUV luminosity. The observed values are indicated by red circles and are offset from their intrinsic value because of extinction. The intrinsic FUV luminosities, derived from the model, are shown as blue diamonds and fall on the solid blue line given by Eq.~(\ref{eq:SFR_FUV}). Region numbers are listed in the figure. A detailed discussion is in the text (\S~\ref{sec:FUVIR}).}
  \label{fig:FUV_SFR}
%\end{center}
\end{figure}

Figure~\ref{fig:FUV_SFR} presents the intrinsic FUV luminosities against the SFRs averaged over the last 100~Myr, for all regions. They are both shown as derived from our SED modeling. The intrinsic and observed FUV luminosities are shown as blue circles and red diamonds, respectively. The blue line represents the theoretical relation given by Eq.~(\ref{eq:SFR_FUV}) and the blue diamonds do not fall exactly on this relation, since they have different SFHs. Not surprisingly, the red circles fall below the blue line for all regions, showing the effect of extinction.

%============================ Figure 07 ============================
%===================================================================

\begin{figure}[b]
%\begin{center}
  \hspace{-0.35in}
  \includegraphics[width=.55\textwidth]{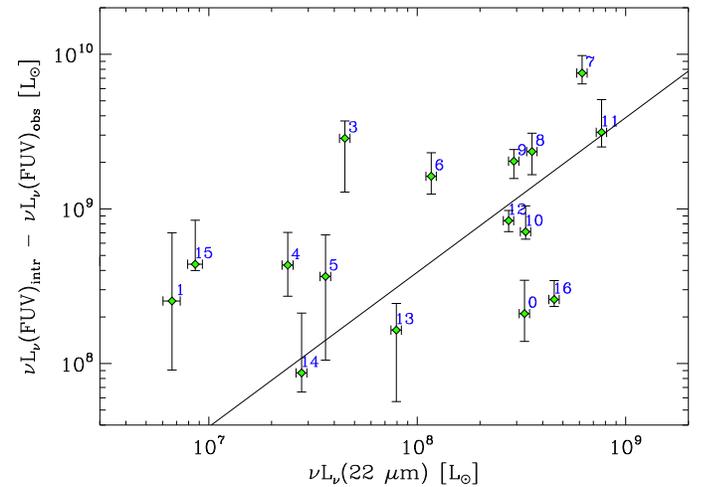}
  \caption{\footnotesize The difference between intrinsic and observed FUV luminosities versus the observed \wise\ 22~\um\ luminosities. The black line shows the expected relation if \acorr\ was constant and equal to 3.89 for all regions. Most points lie above or below this line, showing that a constant value of \acorr\ mostly over- or under-corrects for the effect of the 22~\um\ derived extinction. Region 2 was not detected at 22~\um.}
  \label{fig:dFUV_vs_W4}
%\end{center}
\end{figure}

Figure~\ref{fig:dFUV_vs_W4} shows the difference between intrinsic and observed FUV luminosities versus the observed \wise\ 22~\um\ luminosities. The black line shows the expected relation if \acorr\ was constant, and equal to 3.89 for all regions. Most points lie above or below this line, showing that a constant value of \acorr\ mostly over- or under-corrects for the effect of the 22~\mic\ derived extinction.

%============================ Figure 08 ============================
%===================================================================

\begin{figure}[t]
%\begin{center}
  \hspace{-0.35in}
  \includegraphics[width=.55\textwidth]{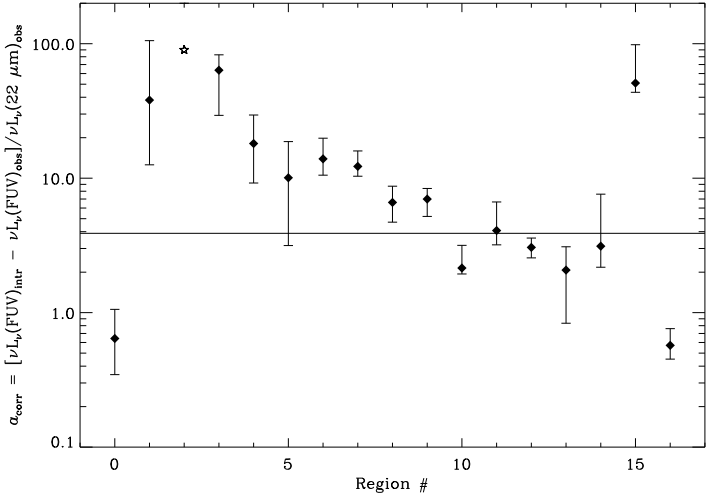}
  \caption{\footnotesize The correlation of \acorr\ with the specific star formation rate, suggesting that the regional variations of \acorr\ arise from variations in the stellar populations that heat the dust. The value of \acorr\ for region 2 (star) is extrapolated from the \wise\ 12~\um.}
  \label{fig:acorr}
%\end{center}
\end{figure}

Figure~\ref{fig:acorr} shows the values of \acorr\ that are required so that the extinction-corrected FUV luminosity equal  the intrinsic FUV luminosities as derived from the SED fittings. The figure shows that a lower value of \acorr\ is needed for Regions 0 and 16, while a much higher value is needed for the youngest regions, especially the outermost Regions as 2, 3 and 15. This implies that a conversion derived from a large sample of galaxies \citep[e.g.][]{Hao2011} overestimates the unobscured SFR for our regions at the galaxies' centers (as Regions 0 and 16) and underestimates it for the outermost regions of the extended arms.

%============================ Figure 09 ============================
%===================================================================

%---------------------------------------------
\subsection{Morphological Effects on \acorr}
%---------------------------------------------
The reason for the variations in the value of \acorr\ for the different regions can be attributed to the different morphology and geometrical relation between the dust and the FUV emitting sources. A case in which all the 22~\mic\ emitting dust is located between the observer and the FUV sources would require a different correction factor compared to a case where the 22~\mic\ emitting dust is equally distributed in front and behind the FUV sources. Obviously, if all the emitting dust would lie behind the source, the introduction of an IR correction would lead to an overestimate in the FUV derived SFR.

On a galactic scale, we expect, from purely statistical reasons, an unbiased distribution of the dust with respect to the FUV emitting sources (e.g. Hao et al. 2011). This should apply to the regions studied here. Since they as large as individual galaxies (10~kpc across), random geometrical effects should average out. Solely on morphological considerations, an average value of \acorr\ may be applicable to galaxies as a whole or to galactic-scale regions in a galaxy, but inadequate in giving the correction for much smaller ($<1~$kpc) spatially-resolved star forming regions, which may have different foreground/background dust structures.

%---------------------------------------------
\subsection{Stellar Lifetime Effects on \acorr, and its Correlation with the sSFR}
%---------------------------------------------
Another source of regional variations in the value of \acorr\ arises from the fact that the FUV and 22~\mic\ emissions sample very different epochs of star formation activities. FUV emission arises predominately from massive, $M > 15$~\msun\ stars, which have lifetimes shorter than 30~Myr. So the FUV emission samples the 30~Myr-averaged SFR. 
The 22~\mic\ emission arises from dust in \hii\ regions as well as stochastically-heated dust residing in the neutral \hi\ gas. The 22~\mic\ emission can therefore originate from dust heated by a large range of stellar masses, as low as $\sim 4$~\msun, which have a main sequence lifetime of 300~Myr. For example, the central 5~kpc of the Condor has copious 22~\mic\ emission and no significant current star formation activity, as reflected by its SED (see Fig.~\ref{fig:SEDs_norm}). Its 22~\mic\ emission is therefore powered by late time stars,  not associated with the star formation of the last 100~Myr. The 22~\mic\ emission therefore samples the SFH of the galaxy that is averaged over a significantly larger epoch than the FUV emission.
Regional variations in \acorr\ may therefore reflect different SFH and different mixes of stellar populations.

\begin{figure}[h]
%\begin{center}
  \hspace{-0.35in}
  \includegraphics[width=.55\textwidth]{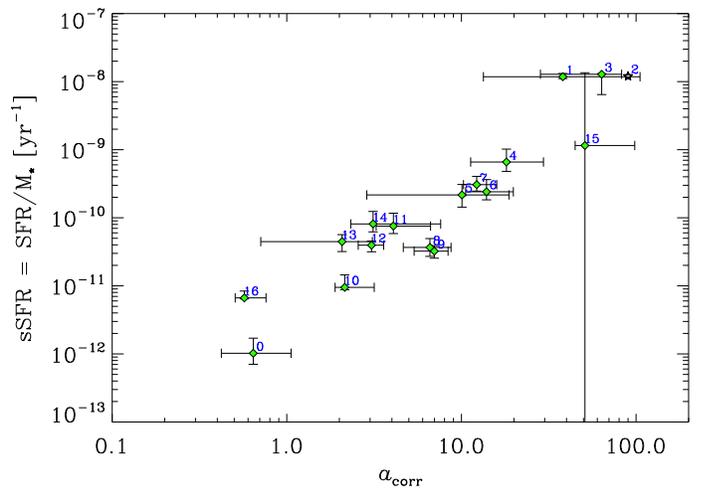}
  \caption{\footnotesize The value of \acorr\ needed for the extinction-corrected FUV luminosity to be equal to the intrinsic one (see Eqs.~\ref{eq:FUV_corr} and \ref{eq:FUV_SFR}) is plotted versus the specific SFR. The value of \acorr\ for region 2 (star) is extrapolated from the \wise\ 12~\um.}
  \label{fig:sSFR100_acorr}
%\end{center}
\end{figure}

To test this idea we plot in Figure~\ref{fig:sSFR100_acorr} the value of \acorr\ as a function of the specific star formation rate, sSFR, defined as the SFR divided by the total stellar mass. The figure shows a trend of increasing \acorr\ as a function of the sSFR. A lower value of the sSFR reflects a overabundance of stars over that expected from a constant SFR at a value of the current epoch. So in regions with a low sSFR, the 22~\mic\ emission samples a longer epoch of star formation activity than in regions with a high sSFR, and a lower value of \acorr\ is needed to correct the observed FUV flux for extinction. This suggests that morphology is not driving the observed variation in \acorr, but the diversity in star formation histories and hence stellar populations.
%============================ Figure 10 ============================
%===================================================================

%%%%%%%%%%%%%%%%%%%%%%%%%%%%%%%%%%%%%%%%%%%%%%%%%%%%%%%%%%%%%%%%%%%
\section{SUMMARY}\label{sec:summary}
%%%%%%%%%%%%%%%%%%%%%%%%%%%%%%%%%%%%%%%%%%%%%%%%%%%%%%%%%%%%%%%%%%%

We have investigated the interacting pair NGC~6872 and IC~4970 (the Condor galaxy and its companion) from the UV to mid-IR, making use of data from \galex, VLT, 2MASS, \spitzer, and \wise. We inspected the \spitzer\ 3.6~\um\ morphology of the Condor galaxy by performing a bulge/bar/disk decomposition. This decomposition showed that both the bar semi-major axis and the disk scale-length of the galaxy are more than double the dimensions of the median values for non-interacting local barred galaxies. The ratio of 1.4, however, is typical in the local universe and befits a giant type-III disk. On top of that, the resemblance of the SEDs of the nucleus and the bar, as well as the lack of a box-peanut structure suggests the bar was probably not formed by this interaction with companion either. In fact, the bar seems to be as old as the central region (Region 0), with an age around 5~Gyr, as derived by our SED fitting. This corroborates the idea that the interaction with the companion IC~4970 affected the Condor less than previously thought and the idea that the Condor galaxy was already a very massive ($M_\star > 10^{11}$\,\msun) and extended galaxy before the interaction.

We have modeled the SFH of the Condor galaxy, NGC~6872, and its interacting companion, IC~4970, on physical scales of the order of 10~kpc. The large size of the galaxy allowed us to sample 17 regions of 10~kpc projected diameter with very distinct spectra, and consequently star formation histories. 

The SEDs of these regions were presented in Table~\ref{tab:SEDs_1} and Figure~\ref{fig:SEDs_norm}. In Figure~\ref{fig:SEDs_norm}, the SEDs are shown normalized by the \spitzer\ 4.5~\um\ measurement, which traces mainly emission from old stars. These normalized SEDs in Figure~\ref{fig:SEDs_norm} show a pronounced trend of bluer regions to the outskirts of the Condor, with the bluest region having an observed FUV/4.5~\um\ flux ratio more than a thousand times higher than the central region of the galaxy. The 22~\micron\ emission, as measured from \wise, demonstrates the presence of dust, and therefore the need to correct the observed SEDs for the effects of extinction.

The UV and mid-IR maps in fact suggest star formation is not concentrated in the bar and central regions of the Condor, as predicted by previous numerical simulations by \cite{Mihos1993}, but mainly in a ring and in the extended tidal arms. The ratio of stellar masses between the Condor and the companion is larger than 12, considerably larger than the one assumed in the previous numerical simulations of 5. The evidence points to the need of examining more closely another scenario proposed by \cite{Mihos1993} in which, prior to the interaction with the companion, the disk of the Condor was not uniform, but a more extended mass distribution. The elliptical galaxy NGC~6876 may have played a larger role in shaping the Condor than previously considered.

The current CO data do not allow us to be conclusive, but the star formation efficiency at the central Region 0 would be remarkably low, with dense molecular gas fraction of $f_{\rm DG}\sim10^{-3}$, were the molecular gas indeed segregated from the \hi\ gas, concentrated at the center of the galaxy. If instead the molecular gas morphology is similar to the one of the \hi\ gas, the central region would have higher star formation efficiency, but a CO map at higher spatial resolution is necessary to confirm this. 

The SEDs could be fit by a single exponential decay, with no need for an additional starburst component. From the SED modeling we find the outermost regions of the northeastern arm (Regions 1 and 2) to be intrinsically the youngest regions. Their best-fit ages ($t_{\rm age}$) around 20~Myr suggest they were assembled after the closest approach of the companion, 130~Myr ago. Regions 3 and 15 do have ages of the order of 100~Myr, what is consistent with being formed during or slightly after closest approach. Regions 5-14, as well the Condor's and its companion's centers (Regions 0 and 16) are all over a gigayear old and were assembled before the interaction with the companion started. If the interaction in fact enhanced SFR in these regions, it was not enough to deviate the SFH of these regions of their long-term, pre-collision SFH, represented by a single exponential decay.

We then investigated the FUV luminosity as a tracer of recent SFR and how the \wise\ 22~\micron\ intensities can be used to estimate the FUV extinction (see Figures \ref{fig:FUV_SFR} and \ref{fig:acorr}). From this analysis we concluded that no single \acorr-correction can be used for all regions and that \acorr\ increases as the sSFR increases. Standard prescriptions (e.g. \citealt{Hao2011}) to correct their observed FUV emission for extinction, as derived from large samples of galaxies, would underestimate the intrinsic FUV emission and consequently the actual SFRs for NGC~6872 taken as whole. However, when viewed in detail, these prescriptions overestimate the intrinsic FUV emission and SFR for the central regions of the Condor and for its companion, and underestimate for outermost regions of the extended arms.

We found that the 22~\um\ ($\sim25$~\um) emission is not always a good proxy for the amount of ``missing" UV emission, since for certain SFHs the 22~\um\ arises from dust heated by the older stellar population.

%%%%%%%%%%%%%%%%%%%%%%%%%%%%%%%%%%%%%%%%%%%%%%%%%%%%%%%%%%%%%%%%%%%%%
%%%%%%%%%%%%%%%%%%%%%%%%%%%%%% ACKNOWLEDGEMENTS %%%%%%%%%%%%%%%%%%%%%
\acknowledgments 
%%%%%%%%%%%%%%%%%%%%%%%%%%%%%%%%%%%%%%%%%%%%%%%%%%%%%%%%%%%%%%%%%%%%%
We thank the anonymous referee for comments that significantly improved the paper. ED acknowledges NASA ADAP proposal NNH11ZDA001N and DFdM was funded by NASA ADAP NNX09AC72G. Based on observations made with the NASA {\it Galaxy Evolution Explorer}. \galex\ is operated for NASA by the California Institute of Technology under NASA contract NAS5-98034. This publication makes use of data products from the Two Micron All Sky Survey, which is a joint project of the University of Massachusetts and the Infrared Processing and Analysis Center/California Institute of Technology, funded by the National Aeronautics and Space Administration and the National Science Foundation.
This work is based on observations made with the Spitzer Space Telescope, obtained from the NASA/IPAC Infrared Science Archive, both of which are operated by the Jet Propulsion Laboratory, California Institute of Technology under a contract with the National Aeronautics and Space Administration.
This publication makes use of data products from the Wide-field Infrared Survey Explorer, which is a joint project of the University of California, Los Angeles, and the Jet Propulsion Laboratory/California Institute of Technology, funded by the National Aeronautics and Space Administration.
%%%%%%%%%%%%%%%%%%%%%%%%%%%%%%%%%%%%%%%%%%%%%%%%%%%%%%%%%%%%%%%%%%%%%

%%%%%%%%%%%%%%%%%%%%%%%%%%%%%%%%%%%%%%%%%%%%%%%%%%%%%%%%%%%%%%%%%%%%%
% Facilities Keywords
%%%%%%%%%%%%%%%%%%%%%%%%%%%%%%%%%%%%%%%%%%%%%%%%%%%%%%%%%%%%%%%%%%%%%
{\it Facilities:}
\facility{\galex}, 
\facility{VLT}, 
\facility{CTIO:2MASS},
\facility{\spitzer},
\facility{\wise}
%%%%%%%%%%%%%%%%%%%%%%%%%%%%%%%%%%%%%%%%%%%%%%%%%%%%%%%%%%%%%%%%%%%%%

%%%%%%%%%%%%%%%%%%%%%%%%%%%%%%%%%%%%%%%%%%%%%%%%%%%%%%%%%%%%%%%%%%%%%
% BIBLIOGRAPHY
%%%%%%%%%%%%%%%%%%%%%%%%%%%%%%%%%%%%%%%%%%%%%%%%%%%%%%%%%%%%%%%%%%%%%
\bibliography{}
%%%%%%%%%%%%%%%%%%%%%%%%%%%%%%%%%%%%%%%%%%%%%%%%%%%%%%%%%%%%%%%%%%%%%

\end{document}